\documentclass[aps,prd,showpacs,nofootinbib,superscriptaddress,amsmath,amssymb]{revtex4-2}
\usepackage{amsfonts}
\usepackage{bm}
\usepackage{verbatim}

\usepackage{graphicx}
\usepackage{graphics}
\usepackage{color}

\newcommand{\revision}[1]{{#1}}

\usepackage{epsfig}

\begin{document}

\title{Momentum Space Topology and Non-Dissipative Currents}

\author{M.A. Zubkov \footnote{On leave of absence from Institute for Theoretical and Experimental Physics, B. Cheremushkinskaya 25, Moscow, 117259, Russia}}
\email{zubkov@itep.ru}
\affiliation{Physics Department, Ariel University, Ariel 40700, Israel}

\author{Z.V.Khaidukov}
\affiliation{Institute for Theoretical and Experimental Physics, B. Cheremushkinskaya 25, Moscow, 117259, Russia}
\affiliation{Moscow Institute of Physics and Technology, 9, Institutskii per., Dolgoprudny, Moscow Region, 141700, Russia}

\author{Ruslan Abramchuk}
\affiliation{Moscow Institute of Physics and Technology, 9, Institutskii per., Dolgoprudny, Moscow Region, 141700, Russia}

\date{\today}

\begin{abstract}
Relativistic heavy ion collisions represent an arena for the probe of various anomalous
transport effects. Those effects, in turn, reveal the correspondence between the solid state
physics and the high energy physics, which share the common formalism of quantum field
theory. It may be shown that for the wide range of {field--theoretic} models, the response of %Please check intended meaning of this sentence has been retained.
various nondissipative currents to the external gauge fields is determined by the momentum
space topological invariants. Thus, the anomalous transport appears to be related to the
investigation of momentum space {topology---the} approach developed earlier mainly in the
condensed matter theory. Within this methodology we analyse systematically the anomalous
transport phenomena, which include, in particular, the anomalous quantum Hall effect, the
chiral separation effect, the chiral magnetic effect, the chiral vortical effect and the rotational Hall effect.
\end{abstract}

\maketitle

\section{Introduction}
\label{intro}
It is expected that the family of the non-dissipative transport effects \cite{Landsteiner:2012kd,semimetal_effects7,Gorbar:2015wya,Miransky:2015ava,Valgushev:2015pjn,Buividovich:2015ara,Buividovich:2014dha,Buividovich:2013hza} will be observed in non-central heavy ion collisions. The fireballs appearing during those collisions exist in the presence of both magnetic field and rotation~\cite{ref:HIC,ref:HIC_,ref:HIC__,ref:HIC___}. It is worth mentioning that the same or similar  effects have been considered for the Dirac and Weyl semimetals \cite{semimetal_effects6,semimetal_effects10,semimetal_effects11,semimetal_effects12,semimetal_effects13,Zyuzin:2012tv,tewary,16}.
The mentioned family consists of the chiral separation effect (CSE) \cite{Metl,Gorbar:2015wya}, the chiral vortical effect (CVE) \cite{Vilenkin}, the anomalous quantum Hall effect (AQHE)  \cite{TTKN,Hall3DTI,Zyuzin:2012tv}, and some other similar phenomena. Besides, the Chiral Magnetic Effect (CME) has been widely discussed   \cite{CME,Kharzeev:2015znc,Kharzeev:2009mf,Kharzeev:2013ffa,SonYamamoto2012,Kharzeev:2009pj}.

The naive “derivations” of the above-mentioned effects using the non-regularized continuous field theory have been presented in some of the above-mentioned publications. Later, the above-mentioned naive derivations were reconsidered (see, for example, \cite{Valgushev:2015pjn,Buividovich:2015ara,Buividovich:2014dha,Buividovich:2013hza}, where the corresponding problems were treated using numerical simulations within the rigorous lattice regularization). The majority of the above-mentioned non-dissipative transport effects were confirmed. However, it was shown, for example, that the equilibrium CME does not exist. The CME may, though, survive as a non-equilibrium kinetic phenomenon-see, for example,  \cite{ZrTe5,Nielsen:1983rb}. In the context of condensed matter theory, the absence of the equilibrium version of the CME was confirmed within the particular model of Weyl \mbox{semimetal \cite{nogo}.} The same conclusion has also been obtained using the {no--go} Bloch theorem \cite{nogo2} (still there is no direct proof of this theorem for the general case of the quantum field theory). %Please check intended meaning of this sentence has been retained.
The other proof was given in~\cite{Z2016_1,Z2016_2}. It is based on the lattice version of Wigner-Weyl formalism \cite{Wigner,star,Weyl,berezin}. This approach also allows one to derive the AQHE \cite{Z2016_1} in $3+1$ D Weyl semimetals, and the CSE \cite{KZ2017} in the lattice regularized quantum field theory. In addition, it allows to derive the axial current of the CVE for the massless fermions at zero temperature.
This technique is reviewed below. Its key point is the intimate relation between momentum space topological invariants and nondissipative currents.

Among the other applications of momentum space topology we mention the investigation of bulk-boundary correspondence for the topological insulators and the question of Fermi surface stability. Besides, this approach is important for the study of fermion zero modes that reside on vortices \cite{Volovik2003,Volovik:2011kg}. In \cite{Z2016_3}, the Momentum space topology of {QCD} was considered while in \cite{Volovik:2016mre} this approach was applied to Standard Model fermions.
%please define where first appears.

\section{Momentum Representation of Lattice Models}

\label{SectWigner}

Following \cite{Z2016_1,Z2016_2} we start here from the lattice model with the  partition function
\begin{equation}
Z = \int D\bar{\psi}D\psi \, {\rm exp}\Big( - \int_{\cal M} \frac{d^D {p}}{|{\cal M}|} \bar{\psi}^T({p}){\cal G}^{-1}({ p})\psi({p}) \Big),\label{Z1}
\end{equation}
where integration is over the fields defined in momentum space. $|{\cal M}|$  is its volume, $D = 4$. $\bar{\psi}$ and $\psi$ are the anticommuting Grassmann variables. $\cal G$ is the one-particle Green function. For the Wilson fermions it is equal to  %\begin{equation}
 ${\cal G}({p}) = - i \Big(\sum_{k}\gamma^{k} g_{k}({p}) - i m({p})\Big)^{-1}$,
 with the Euclidean Dirac matrices $\gamma^k$.
while $g_k({p})$ and $m({p})$  ($k = 1,2,3,4$) are
$g_k({p}) = {\rm sin}\,
p_k, \quad m({p}) = m^{(0)} +
\sum_{a=1,2,3,4} (1 - {\rm cos}\, p_a)$.
In coordinate space we have
$\psi({r}) = \int_{\cal M} \frac{d^D {p}}{|{\cal M}|} e^{i {p}{r}} \psi({p})$.
 This expression gives the values of the fermionic field defined on the lattice sites. However, it also defines the values of this field for the continuous coordinates $r$. Then the partition function receives the form
\begin{equation}
Z = \int D\bar{\psi}D\psi \, {\rm exp}\Big( - \sum_{{r}_n} \bar{\psi}^T({r}_n){\cal G}^{-1}(-i\partial_{r})\psi({ r})\Big|_{{r}={r}_n} \Big).\label{Z2}
\end{equation}

The sum in this expression is over the lattice sides ${r}_n$. At the same time $-i\partial_{r}$ is applied to the defined above function $\psi({r})$ of real-valued coordinates.

\section{External Gauge Field in Momentum {Space: Chiral} Magnetic Effect}  %Please check intended meaning of this sentence has been retained.
\label{SectCME}

The CME was originally understood as the equilibrium phenomenon---the appearance of electric current in the presence of magnetic field in the system of massless fermions with chiral imbalance. The latter has been introduced with the aid of the chiral chemical potential. The early derivations of the CME \cite{Kharzeev:2009pj} dealt with the non-regularized continuum field theory, in which the chiral chemical potential was considered on the same grounds as the ordinary one. Namely, it was supposed that it results in the appearance of the two distinct Fermi surfaces for the left-handed and the right-handed fermions. The more rigorous consideration discussed here relies on the lattice regularized version of the theory. If the chiral chemical potential is introduced to this theory, the Fermi surface does not appear. Instead, the fermions become gapped, and the method discussed below of the electric current calculation may be applied.

Let us add to the lattice system described above the external gauge field potential $A(x)$. The slightly modified partition function of the resulting model receives the form \cite{Z2016_1,Z2016_2}
\begin{eqnarray}
Z = \int D\bar{\psi}D\psi \, {\rm exp}\Big( -  \int_{\cal M} \frac{d^D {p}}{|{\cal M}|} \bar{\psi}^T({p})\hat{\cal Q}(i{\partial}_{p},{p})\psi({p}) \Big).\label{Z4}
\end{eqnarray}

It differs from the precise expression by terms that disappear in continuum limit. In the above expression
%\begin{equation}
$\hat{\cal Q} = {\cal G}^{-1}({p} - {A}(i{\partial}^{}_{p}))$
%\label{calQM}\end{equation}
while operator ${A}(i\partial_{p})$ represents the gauge field ${A}({r})$ as a function of coordinates ${r}$. Instead of the coordinate ${r}$, we  substitute the differential operator $i\partial_{p}$. The products of the components of ${p} - {A}(i{\partial}^{}_{p})$ are symmetrized \cite{Z2016_1}.

%\section{Electric current}

In order to calculate Electric current we consider variation of external Electromagnetic field. The corresponding variation of effective action gives
\begin{eqnarray}
j^k({R}) &=& -\int_{\cal M} \frac{d^D {p}}{(2\pi)^D} \,  {\rm Tr} \, \tilde{G}({R},{p}) \frac{\partial}{\partial p_k}\Big[\tilde{G}^{(0)}({R},{p})\Big]^{-1}.\label{j423}
\end{eqnarray}

This expression contains Wigner transformation
%\begin{equation}
$ \tilde{G}({R},{p}) = \sum_{{r}={r}_n} e^{-i {p} {r}} G({R}+{r}/2,{R}-{r}/2)$
 %\label{Wl2} \end{equation}
of the Green function
\begin{eqnarray}
G({r}_1,{r}_2)&=& -\frac{1}{Z}\int D\bar{\Psi}D\Psi \,\bar{\Psi}({r}_2)\Psi({r}_1) {\rm exp}\Big(-\sum_{{ r}_n}\Big[ \bar{\Psi}({r}_n)\Big[{\cal G}^{-1}(-i\partial_{r}   - {A}({r}))\Psi({r})\Big]_{{r}={r}_n}\Big]\Big),\nonumber
\end{eqnarray}
while $
%\begin{eqnarray}
%&&
\tilde G^{(0)}({R},{p})  = {\cal G}({p}-{A}({R}))\label{Q0}
%\end{eqnarray}
$.

The linear response of electric current to electromagnetic field has the form \begin{eqnarray}
j^{(1)k}({R})  &= &- \frac{1}{4\pi^2}\epsilon^{ijkl} {\cal M}_{l} A_{ij} ({R}), \quad \label{calM}
{\cal M}_l = \int_{} \,{\rm Tr}\, \nu_{l} \,d^4p \label{Ml}, \\ \nu_{l} & = &  - \frac{i}{3!\,8\pi^2}\,\epsilon_{ijkl}\, \Big[  {\cal G} \frac{\partial {\cal G}^{-1}}{\partial p_i} \frac{\partial  {\cal G}}{\partial p_j} \frac{\partial  {\cal G}^{-1}}{\partial p_k} \Big].  \label{nuG}
\end{eqnarray}

All components of $\cal M$ are the topological invariants. When the system is changed smoothly, they are not changed unless the phase transition is encountered.

In the presence of the chiral chemical potential $\mu_5$, the Green function receives the form \cite{Z2016_2}:
 \begin{equation}
 {\cal G}^{}({\bf p}) = \Big(\sum_{k}\gamma^{k} g_{k}({\bf p}) + i\gamma^4 \gamma^5 \mu_5 - i m({\bf p})\Big)^{-1}.\label{G2}
 \end{equation}

Functions $g_k({\bf p})$ and $m({\bf p})$ are {real-valued}, $k = 1,2,3,4$.  At $\mu_5=0$ there is the pole of $\cal G$. This means the presence of  massless excitation. As mentioned above, for the nonzero $\mu_5$, the fermions become gapped.  %Please check intended meaning of this sentence has been retained.

${\cal G}$ of Equation (\ref{G2}) is to be substituted to Equation (\ref{nuG}). Thus, one obtains an expression for the linear response of electric current to  external magnetic field. If there is the CME, this response would be proportional to the chiral chemical potential. However, we prove that it is equal to zero identically. The sketch of the proof is as follows.
For the lattice regularization of the QFT with Wilson fermions, we modify the system continuously starting from the state with  $\mu_5 \ne 0$ and zero mass. This way we move it to the state with $\mu_5 = 0$ and nonzero mass. The value of ${\cal M}_4$ (responsible for the CME), being the topological invariant, is not changed during this modification. For the latter system ${\cal M}_4 =0$. This proves the absence of the equilibrium version of CME \cite{Z2016_2}.

It is worth mentioning that the non-equilibrium version of the CME still may exist. Negative magnetoresistance in Dirac semimetals (see, for example, \cite{ZrTe5}) is often associated with such a version of the CME. Besides, we would like to note the following paper on the {non-equilibrium} dynamics of the CME \cite{CMEchina}.  %Please check intended meaning of this sentence has been retained.
\section{Anomalous Quantum Hall Effect}
\label{SectHall3d}

The AQHE is the appearance of the electric current orthogonal to the applied electric field without the external magnetic field. It is generally believed that this effect is related to the internal topology of the material. In the non-interacting two-dimensional systems the AQHE conductivity is proportional to the so-called TKNN invariant \cite{TTKN} composed of the Berry curvature. In the interacting $2$D systems, this conductivity is given by the invariant composed of the Green functions \cite{Volovik2003}, which is reduced to the TKNN invariant in the non-interacting case. We extend the relation between the topological invariants and the AQHE conductivity to the three dimensional systems.

%\subsection{General case}

Namely, in the $3+1$ D systems, the Hall current is given by \cite{Z2016_1}
\begin{equation}
{j}^k_{Hall} = -\frac{1}{4\pi^2}\,{\cal M}^\prime_l\,\epsilon^{jkl}E_j,\label{HALLj3d}
\end{equation}
where
\begin{eqnarray}
{\cal M}^\prime_l &=&  \frac{1}{3!\,4\pi^2}\,\epsilon_{ijkl}\,\int_{} \,\,d^4p\,{\rm Tr} \Big[  {\cal G} \frac{\partial {\cal G}^{-1}}{\partial p_i} \frac{\partial  {\cal G}}{\partial p_j} \frac{\partial  {\cal G}^{-1}}{\partial p_k} \Big].  \label{nuGHall}
\end{eqnarray}

For the condensed matter system with the Green function  ${\cal G}^{-1} = i \omega - \hat{H}({\bf p})$ (where $\hat{H}$ is the one-particle Hamiltonian), the topological invariant ${\cal M}^{\prime}_l$ with $l\ne 4$ may be expressed as follows
%\begin{eqnarray}
${\cal M}^\prime_l =  \frac{\epsilon^{ijl}}{4\pi}\sum_{\rm occupied}\, \int d^3p\, {\cal F}_{ij}$.
%\end{eqnarray}
Here the sum is over the occupied branches of spectrum, while $ {\cal F}_{ij}$ is the Berry curvature.

From Equation (\ref{HALLj3d}) it follows that in Weyl semimetals, each pair of Weyl fermions contributes to the AQHE current:
\begin{equation}
{j}^k_{Hall} = \frac{\beta}{2\pi^2}\,\epsilon^{jk3}E_j\label{HALLj3dp}.
\end{equation}

It is assumed here that the given pair of the Weyl points is situated along the $z$ axis. $\beta$ is the distance between them. This result confirms the expressions obtained earlier using the non-regularized naive theory.

In \cite{Z2016_1}, the method was proposed that allows to calculate the  topological invariant responsible for the AQHE in the $3$D materials for the wide class of the systems with $2\times 2$ and $4\times 4$ Green functions. In the same paper, the topological insulators of certain types were discussed, in which there is the AQHE, and the resulting Hall current is proportional to one of the vectors of the reciprocal lattice. The similar models were also considered earlier using the other methods in \cite{Klinkhamer:2004hg}, where the authors draw the same conclusion about the existence of the AQHE. In \cite{Z2016_1}, using the Wigner-Weyl formalism, it has also been shown that the insulator state, which admits the AQHE, is accompanied by the surface Fermi lines. In the $3+1$ D Weyl semimetals, those Fermi lines are reduced to the Fermi arcs connecting the bulk Weyl points in accordance to the results obtained earlier via the more \mbox{conventional methods.}

\section{Chiral Separation Effect}\label{sec5}

The CSE was proposed in \cite{Metl}. It is the appearance of the axial current in the presence of the external magnetic field in the fermionic system with free charge carriers caused by chemical potential or temperature. The early derivations of this effect were very much similar to the early derivations of the equilibrium CME. Therefore, the analysis reported in Section \ref{SectCME} prompts that the existence of the CSE should be checked carefully. The careful consideration within the lattice regularized theory confirms the existence of this effect and relates it in the massless case for vanishing temperature to the topological invariant in momentum space responsible for the stability of the Fermi points.

For the theory in lattice regularization, there are many different definitions of axial current that give $\langle \bar{\psi} \gamma^\mu \gamma^5 \psi\rangle$ in the continuum limit. The definition proposed in \cite{KZ2017} reads
\begin{eqnarray}
j^{5k}({R}) &=& \int_{\cal M} \frac{d^D {p}}{(2\pi)^D} \,  {\rm Tr} \,\gamma^5\,  \tilde{G}({R},{p}) \frac{\partial}{\partial p_k}\Big[\tilde{G}^{(0)}({R},{p})\Big]^{-1},\label{j423}
\end{eqnarray}
where
\begin{eqnarray}
&\tilde G^{(0)}({R},{p})  = {\cal G}({p}-{A}({R}))\label{Q01}.
\end{eqnarray}

We regularize the infrared divergencies via finite temperature.
Correspondingly, the momenta are given by
\begin{equation}\label{disc}
p_i \in (0, 2\pi);\, p_4=\frac{2\pi}{N_t }(n_4+1/2).
\end{equation}

Here $i=1,2,3$ and $n_4=0,...,N_t-1$. In lattice units the temperature is $T = 1/N_t $. The Matsubara frequencies are $p_4 = \omega_{n}=2\pi T (n+1/2)$, where $n= 0, 1, ..., N_t-1$. The axial current is given by
\begin{eqnarray}
j^{5k}&=&-\frac{i}{2}T\sum_{n=0}^{N_t-1}\int \frac{d^3p}{(2\pi)^3}{\rm Tr}\, \gamma^5 ({\cal G}(\omega_{n},\textbf{p})\partial_{p_{i}}{\cal G}^{-1}(\omega_{n},\textbf{p}) \partial_{p_{j}}{\cal G}(\omega_{n},\textbf{p})\partial_{p_{k}}{\cal G}^{-1}(\omega_{n},\textbf{p}))F_{ij}.\label{calN}
\end{eqnarray}

%\section{The linear response of the chiral current to chemical potential and to external magnetic field}

The chemical potential is added via modification $\omega_{n} \to \omega_{n}-i\mu$. Suppose that $\{\gamma^5, {\cal G}\}=0$ are close to the poles of the Green function. Then \cite{KZ2017}
\begin{equation}
j^{5k}= \frac{{\cal N}\,\epsilon^{ijk}}{4\pi^2} F_{ij} \mu\label{jmuH},
\end{equation}
at $T\to 0$. Here we encounter the following topological invariant
\begin{eqnarray}
{\cal N}&=&\frac{\epsilon_{ijkl}}{12}\int_{\Sigma}\frac{d \sigma^l}{(2\pi)^2}{\rm Tr}\, \gamma^5 {\cal G}(\omega_{},\textbf{p})\partial_{p_{i}}{\cal G}^{-1}(\omega_{},\textbf{p})\partial_{p_{j}}{\cal G}(\omega_{},\textbf{p}) \partial_{p_{k}}{\cal G}^{-1}(\omega_{},\textbf{p}).\label{calN1}
\end{eqnarray}

$\Sigma$ is the hypersurface embracing the poles of the Green function. It is supposed that the volume of this hypersurface is infinitely small.
For the lattice model, which gives rise to one Dirac fermion in the continuum limit, we obtain
%\begin{equation}
${\cal N}= 1$.
%\end{equation}
This confirms the previously-obtained expression for the CSE.

In \cite{KZ2017}, it has been shown that the naive continuum derivation of the CSE is ambiguous, and the ambiguities are related to the difference in the  order of the integration over the $3$-momenta and the summation over the  Matsubara frequencies. If the sum is performed first, and the integral is calculated later on, then there are no  divergencies while the expression of \cite{Metl} for the CSE is reproduced.  At the same time, if the order is inverse, then the resulting expression is not well defined. This consideration demonstrates the necessity to consider the rigorous ultraviolet regularization. However, the analysis reported above demonstrates that the conventional expression for the CSE does appear for the lattice regularized systems of massless Dirac fermions.

\section{Rotational Hall Effect}
\label{SecFree}

In \cite{RotHall}, the new non-dissipative transport effect was proposed---the rotational Hall effect (RHE). Let us suppose that the fermionic system is  rotated, and the angular velocity \(\Omega\) depends on the distance to the rotation axis $r$. The system is supposed to be in the steady state with the chemical potential \(\mu\) depending on $r$. Interactions are neglected. The linear velocity of substance $\Omega r$ remains smaller than the speed of light (that is $1$ in our units). It has been shown in \cite{RotHall}
that if the electric field orthogonal to $\Omega$ is added, then the electric current appears orthogonal both to the electric field and to the axis of rotation.

In \cite{chiralhydro}, it has been proposed that the action corresponding to a macroscopic 4-velocity $u_\mu$ \mbox{is given by}
\begin{equation}
        S = \int d^4x \bar{\psi}(\gamma^\mu(i\partial_\mu+\mu u_\mu)-M)\psi.\label{action}
\end{equation}

For the case of rotation, we have
    \begin{equation}
        u^\mu = \gamma(r)(1,-\Omega(r) y,\Omega(r) x,0)^T,\quad \gamma(r)=\frac{1}{\sqrt{1-\Omega^2(r)r^2}}.
    \end{equation}

Thus, the emergent Abelian gauge field potential $A_\mu = - \mu u_\mu$ appears.

Let us consider the case when there is the electric field ${\bf E}_{ext}$  orthogonal to the axis of rotation. We suppose, that \revision{$M^2 \gg |2\mu\Omega| \gg |E_{ext}|$},  $\mu_{lab} = \mu(r) \gamma(r) = const$. Then the emergent gauge field is
\begin{equation}
A^\mu = (-\mu_{lab}-E_{ext}x,\mu_{lab} \Omega(r) y,-\mu_{lab}\Omega(r) x,0)^T. \label{Aeff}
\end{equation}

It results in magnetic field $\mathbf{B} = (0,0, -2\mu_{lab}\Omega(r)  [1+\frac{1}{2} \frac{d {\rm log}\,\Omega(r) }{d\,{\rm log}\,r}]) $ and  electric field $\mathbf{E} = (E_{ext},0,0)$.
We also assume that $l_B M \gg 1,\quad \frac{d l_B}{d r}\sim \frac{d (1/\sqrt{\mu_{lab}\Omega})}{dr} \ll 1 \label{magnlength}
$.

The states corresponding to Landau Levels drift along the equipotential lines that are parallel to axis $y$. The drift velocity is equal to
\begin{equation}
{\bf V} = \frac{{\bf E}_{ext}\times {\bf \Omega}}{2 \mu_{lab}  \Omega^2}\,.
\end{equation}

It is orthogonal to the electric field and to the axis of rotation.

Using the above-mentioned approach, we are able to consider the quark-gluon plasma with several types of quarks that correspond to chemical potentials $\mu_f$. At high temperatures, the Hard Thermal Loop approximation  \cite{Andersen:2012wr} allows to calculate the  thermodynamic potential $\Omega^{(HTL)QCD}$, which gives electric current directed along the mentioned drift velocity
\begin{equation}
{\bf j}^{(HTL)QCD} =   \sum_f \frac{\partial \Omega^{(HTL)QCD}}{\partial \mu^2_f}Q^2_f\frac{{\bf E}_{ext}\times {\bf \Omega}}{ \Omega^2}.
\end{equation}

It is worth mentioning that at sufficiently small values of temperature, the  nonperturbative contributions to this current are important. However, this is out of the scope of the present paper.

\section{Chiral Vortical Effect}
\label{SectOM}

CVE was proposed in \cite{Vilenkin}. This is the appearance of the axial current along the axis of rotation in the presence of chemical potential or temperature. Since rotation in many aspects is similar to the presence of magnetic field, we are able to relate the CVE with the CSE. As a result, the corresponding conductivity also appears to be proportional to the topological invariant in momentum space for the case of vanishing mass and small temperature.

In this section, we start from the description of rotation given above in  Section \ref{SecFree}.  Our consideration follows \cite{ACVE}.
In this approach, the Dirac equation that describes fermions in the presence of rotation with angular velocity $\Omega$ is identical to the Dirac equation in the presence of emergent magnetic field directed along the axis of rotation.
Its value is given by $-2\mu \Omega$.

We assume, that the fermions are massless. At this point we are able to use the results of \cite{KZ2017} on the Chiral Separation effect. Namely, the axial current along the rotation axis appears given by Equations (\ref{jmuH}) and (\ref{calN1}).

We obtain, therefore, the Chiral Vortical Effect with the axial current given by
 \begin{equation}
j^{5k}= \frac{{\cal N} \epsilon^{12k}}{2\pi^2}  \mu^2 \Omega \label{jmuH_}.
\end{equation}

Topological invariant ${\cal N}$ is expressed by Equation (\ref{calN1}).

Notice that at finite temperature, the given description of rotation is not exhaustive. Namely, rotation of the quasiparticles excited due to the finite temperature cannot be described by emergent gauge field $-\mu u_\nu$. Therefore, we use another rotation description, in which the following term is added to the action (instead of the emergent gauge field $-\mu u_\nu$)
$$
\delta S = \Omega \int d t \omega_{ij} M^{ij}.
$$

Here $\omega_{ij} = \epsilon_{03ij}$ while momentum tensor is given by:
$$
M^{ij} = \frac{1}{2}\int d^3 x \bar{\psi}\Big(\gamma^0\{ x^i ,\hat{P}^j\} -\gamma^0 \{x^j, \hat{P}^i\} + \{\gamma^0 , \frac{1}{2}\Sigma^{ij}\}\Big)\psi.
$$

This leads to equation
 \begin{equation}
        \Big[\gamma^0\mu  + \gamma^0\Omega \frac{\epsilon_{03ij}}{2}\Big(i\{ x^i ,\partial^j\} -i \{x^j, \partial^i\} +  \Sigma^{ij}\Big) + i \gamma^\mu \partial_\mu-M\Big]\psi=0. \label{direq2}
     \end{equation}

Comparing this expression with {Equation (2.17)} of \cite{chernodub}, we see that the presented description of rotation is equivalent to the description of the system in rotated reference frame (see also \cite{Ambrus:2014uqa}).%please check which equation this is.

In order to avoid the appearance of the linear velocity exceeding the speed of light, we have to consider the system inside the cylinder. At the boundary of the cylinder, the {MIT} bag \mbox{conditions are applied}%please define it where first appears.
\begin{eqnarray}
  &&  0=(i\gamma^\mu n_\mu -1)\psi\rvert_{r=R},\label{mitbc}
\end{eqnarray}
 where $n_\mu = (0, \frac{\bf r}{r}, 0)$ is directed along the radius of  the cylinder. These boundary conditions give vanishing electric current across the cylinder surface.

Here we rely on the following definition of the axial current
\begin{equation}
    j^5_z = \bar{\psi}\gamma^5\gamma^3\psi.
\end{equation}

Notice that the definition of Equation (\ref{j423}) is {reduced to it} in the continuum limit.  The Fermi distribution is given by %reduced to what? Please confirm.
%\begin{equation}
 $n(w,j_z) = \frac{1}{e^{\beta(w - \mu - \Omega j_z)}+1}$.
%\end{equation}
Here $j_z$ is the $z$ component of the total angular momentum, while
$w = E +\mu +\Omega j_z$, and $E$ is the energy counted from the Fermi level.

Subtracting the contribution of vacuum to rotation we
 take
\begin{equation}
    n(w,j_z) = \frac{{\rm sign}\,(w )}{e^{\beta(w - \mu - \Omega j_z){\rm sign}\,(w)}+1}.
\end{equation}
and obtain the following axial current
\begin{equation}
    \langle j^5_z(r)\rangle_\beta = \sum_{k,q,{\rm sign}\,(w),j_z} n(w,j_z) \bar{\psi}_{k,q,{\rm sign}\,(w),j_z}\gamma_5\gamma_3\psi_{k,q,{\rm sign}\,(w),j_z}. \label{j5avg}
\end{equation}

Here $\psi_{k,q,{\rm sign}\,(w),j_z}$ is the eigenfunction corresponding to the quantum numbers: angular momentum $j_z$, $z$-component of momentum $k$, radial number $q$, while $\bar{\psi} = (\psi^*)^T \gamma^0$.

The axial current of Equation (\ref{j5avg}) is calculated numerically.
Our results on the dependence of the axial current on $\mu$ at the rotation axis ensure that the data are fitted well by Equation (\ref{jmuH_}) if $R$ is large enough and consequently $T\sim1/R$ is small enough, while $T$ remains larger than $\Omega$. Under these conditions, the CVE conductivity obtained  for the rotation description of Equation (\ref{direq2}) is given by the topological invariant in momentum space.
This ensures that the coefficient at the term $\sim \mu^2$ in the CVE is topologically protected at vanishing temperature. If we change the system smoothly, this coefficient is not changed. The introduction of interactions is also a kind of the smooth deformation of the system and it also cannot renormalize this coefficient. At large temperature there is not such a remarkable correspondence between the two above-mentioned approaches to the definition of rotation, and this prompts that the corresponding term proportional to $T^2$ in the CVE current of \cite{Metl}  may be modified due to the interactions in accordance with the conclusions of \cite{Corr1,Corr2}.

\section{Conclusions}
\label{sec-1}

Above we reviewed the application of momentum space topology to the analysis of anomalous transport reported earlier in our papers \cite{Z2016_1,Z2016_2,KZ2017,ACVE,RotHall}. This methodology works in the lattice regularized relativistic quantum field theory, and also may be applied to the solid state physics. It appears that the conductivities corresponding to the considered non-dissipative transport phenomena in many cases are proportional to the topological invariants in momentum space. In this way we prove the absence of the equilibrium CME, and describe the axial current of the CSE, and the CVE (for zero temperature). Besides, we describe the AQHE in the $3+1$ D systems and introduce the new {non-dissipative} rotational Hall effect. %Please check intended meaning has been retained.
It is worth mentioning that for the nonzero fermion mass, the corrections to the CSE appear that break the discussed relation to the topological invariants in momentum space (see, for example, \cite{Gorbar:2015wya,Miransky:2015ava}). This does not contradict our results because we relate the CSE conductivity to the topological invariant in the massless case only. The same refers also to the mass and temperature corrections to the CVE, discussed in \cite{ACVE,Corr1,Corr2}.

The mentioned phenomena may be relevant for the description of the fireballs that appear during the non-central heavy ion collisions. Those fireballs are widely believed to be in the semi-equilibrium steady state. In addition, the kinematics of the heavy ion collisions ensures the presence of the strong magnetic field. Besides, the fireballs are rotated. The temperatures of the appeared fireballs are so large that the quark matter is supposed to be in the quark---gluon plasma phase---the state of matter with almost massless quarks.  Their current masses remain, however, for the $u$ and $d$ quarks these masses are several orders of magnitude smaller than their constituent masses inside protons and neutrons. This allows to study the almost massless fermions in the presence of both magnetic field and rotation. This is a suitable arena for the investigation of the CSE, the CVE, and the rotational Hall effect (RHE).

The non-dissipative transport effects are also relevant for the physics of the electronic quasiparticles in solids. The AQHE is typical for the three-dimensional Weyl semimetals that were predicted and discovered recently, and also for a certain class of topological insulators, which is not yet discovered experimentally. The CSE may also be relevant for the solid state physics. In particular, this effect is to be present in the Weyl/Dirac semimetals, where the emergent Dirac quasiparticles are massless. It is worth mentioning that the methods for the observation of axial current in the Weyl/Dirac semimetals are not yet developed and this complicates the experimental observation of the CSE in those materials. Finally, we would like to notice the reach perspectives for the investigation of the anomalous transport in {Helium--3} superfluid (for details see \cite{Volovik2003}). %Please check intended meaning has been retained.

%%%%%%%%%%%%%%%%%%%%%%%%%%%%%%%%%%%%%%%%%%

%%%%%%%%%%%%%%%%%%%%%%%%%%%%%%%%%%%%%%%%%%
\vspace{6pt}

%%%%%%%%%%%%%%%%%%%%%%%%%%%%%%%%%%%%%%%%%%
%% optional
%\supplementary{The following are available online at \linksupplementary{s1}, Figure S1: title, Table S1: title, Video S1: title.}

% \supplementary{The following are available at \linksupplementary, Figure S1: title, Table S1: title, Video S1: title. A supporting video article is available at doi: link.}

%%%%%%%%%%%%%%%%%%%%%%%%%%%%%%%%%%%%%%%%%%
{The work presented in Sections \ref{SectWigner}--\ref{SectHall3d} and \ref{SecFree} has been performed by M.Z. The work presented in Section \ref{sec5} was done by Z.K. and M.Z. The work presented in Section \ref{SectOM} was performed by R.A., Z.K. and M.Z. The general supervision of the project: M.Z.}

%%%%%%%%%%%%%%%%%%%%%%%%%%%%%%%%%%%%%%%%%%
{The work of Z.K. was supported by Russian Science Foundation Grant No 16-12-10059.}

%%%%%%%%%%%%%%%%%%%%%%%%%%%%%%%%%%%%%%%%%%
\acknowledgments{M.Z. kindly acknowledges numerous discussions with Volovik, G.E. M.Z. and Z.K. are grateful to Chernodub, M.N. for useful discussions. R.A. is grateful for useful comments to Zakharov, V.I.  }

%%%%%%%%%%%%%%%%%%%%%%%%%%%%%%%%%%%%%%%%%%
\vspace{6pt}
%%%%%%%%%%%%%%%%%%%%%%%%%%%%%%%%%%%%%%%%%%
%% optional
{The following abbreviations are used in this manuscript:\\
\noindent
\begin{tabular}{@{}ll}
CME & Chiral Magnetic Effect\\
CSE & Chiral Separation Effect\\
AQHE & Anomalous Quantum Hall Effect\\
CVE & Chiral Vortical Effect \\
RHE & Rotational Hall Effect
\end{tabular}}

%%%%%%%%%%%%%%%%%%%%%%%%%%%%%%%%%%%%%%%%%%
%% optional
%\appendixtitles{no} %Leave argument "no" if all appendix headings stay EMPTY (then no dot is printed after "Appendix A"). If the appendix sections contain a heading then change the argument to "yes".
%\appendixsections{multiple} %Leave argument "multiple" if there are multiple sections. Then a counter is printed ("Appendix A"). If there is only one appendix section then change the argument to "one" and no counter is printed ("Appendix").
%\appendix
%\section{}
%\subsection{}


%=====================================
% References, variant A: internal bibliography
%=====================================
%\reftitle{References}

\end{document}